\documentclass[10pt,conference]{IEEEtran}
\IEEEoverridecommandlockouts
\usepackage{cite}
\usepackage{amsmath,amssymb,amsfonts}
\usepackage{algorithmic}
\usepackage{graphicx}
\usepackage{textcomp}
\usepackage{xcolor}

\usepackage{listings}
\usepackage{array} 
\usepackage{ragged2e}
\usepackage{enumitem}
\usepackage{todonotes}
\usepackage{url}
\usepackage{wrapfig}
\usepackage{subcaption}
\usepackage{caption}
\usepackage{multirow}
\usepackage{flushend}
\usepackage{tikz}
\usepackage{censor}
\usepackage[most]{tcolorbox}
\usepackage{soul}

\usepackage{textcomp}
\usepackage{amsmath}
\usepackage{array, multirow, bigdelim, makecell, booktabs}

\tcbuselibrary{theorems}
\newtcolorbox{mybox}[1]{colback=white!5!white,colframe=gray!45!white, title =#1,coltitle=black!20!black}

\newcommand{\fig}[1]{Fig.~\ref{fig:#1}}

\colorlet{punct}{red!60!black}
\definecolor{background}{HTML}{EEEEEE}
\definecolor{delim}{RGB}{20,105,176}
\colorlet{numb}{magenta!60!black}

\definecolor{plantucolor0000}{RGB}{24,24,24}
\definecolor{plantucolor0001}{RGB}{241,241,241}
\definecolor{plantucolor0002}{RGB}{0,0,0}

\lstdefinelanguage{json}{
    basicstyle=\normalfont\ttfamily,
    numbers=left,
    numberstyle=\scriptsize,
    stepnumber=1,
    numbersep=8pt,
    showstringspaces=false,
    breaklines=true,
    frame=lines,
    backgroundcolor=\color{background},
    literate=
     *{0}{{{\color{numb}0}}}{1}
      {1}{{{\color{numb}1}}}{1}
      {2}{{{\color{numb}2}}}{1}
      {3}{{{\color{numb}3}}}{1}
      {4}{{{\color{numb}4}}}{1}
      {5}{{{\color{numb}5}}}{1}
      {6}{{{\color{numb}6}}}{1}
      {7}{{{\color{numb}7}}}{1}
      {8}{{{\color{numb}8}}}{1}
      {9}{{{\color{numb}9}}}{1}
      {:}{{{\color{punct}{:}}}}{1}
      {,}{{{\color{punct}{,}}}}{1}
      {\{}{{{\color{delim}{\{}}}}{1}
      {\}}{{{\color{delim}{\}}}}}{1}
      {[}{{{\color{delim}{[}}}}{1}
      {]}{{{\color{delim}{]}}}}{1},
}

\definecolor{eclipseStrings}{RGB}{42,0.0,255}
\definecolor{eclipseKeywords}{RGB}{127,0,85}
\colorlet{numb}{magenta!60!black}

\newcounter{theo}[section]
\renewcommand{\thetheo}{\arabic{section}.\arabic{theo}}

\definecolor{airforceblue}{rgb}{0.36, 0.54, 0.66}
\definecolor{amber}{rgb}{1.0, 0.75, 0.0}
\definecolor{antiquebrass}{rgb}{0.8, 0.58, 0.46}
\definecolor{bleudefrance}{rgb}{0.19, 0.55, 0.91}
	\definecolor{bittersweet}{rgb}{1.0, 0.44, 0.37}
	\definecolor{bondiblue}{rgb}{0.0, 0.58, 0.71}

\colorlet{mylinkcolor}{bondiblue}
\colorlet{mycitecolor}{bittersweet}
\colorlet{myurlcolor}{bondiblue}

\makeatletter
\newcommand\footnoteref[1]{\protected@xdef\@thefnmark{\ref{#1}}\@footnotemark}
\makeatother

\colorlet{punct}{red!60!black}
\definecolor{background}{HTML}{EEEEEE}
\definecolor{delim}{RGB}{20,105,176}
\colorlet{numb}{magenta!60!black}

\definecolor{gray50}{gray}{.5}
\definecolor{gray40}{gray}{.6}
\definecolor{gray30}{gray}{.7}
\definecolor{gray20}{gray}{.8}
\definecolor{gray10}{gray}{.9}
\definecolor{gray05}{gray}{.95}

\newlength\Linewidth
\def\findlength{\setlength\Linewidth\linewidth
    \addtolength\Linewidth{-4\fboxrule}
    \addtolength\Linewidth{-3\fboxsep}
}

\newenvironment{rqbox}{\par\begingroup
	\setlength{\fboxsep}{5pt}\findlength
	\setbox0=\vbox\bgroup\noindent
	\hsize=0.95\linewidth
	\begin{minipage}{0.95\linewidth}\normalsize}
	{\end{minipage}\egroup
	\textcolor{gray20}{\fboxsep1.5pt\fbox
		{\fboxsep5pt\colorbox{gray05}{\normalcolor\box0}}}
	\endgroup\par\noindent
	\normalcolor\ignorespacesafterend}

\def\BibTeX{{\rm B\kern-.05em{\sc i\kern-.025em b}\kern-.08em
    T\kern-.1667em\lower.7ex\hbox{E}\kern-.125emX}}

\begin{document}



\title{Fostering Microservice Maintainability Assurance through a Comprehensive Framework}

\author{\IEEEauthorblockN{Amr S. Abdelfattah}
\IEEEauthorblockA{Computer Science Department, Baylor University \\
Waco, TX, USA \\
amr\_elsayed1@baylor.edu\\
Supervisors: Tomas Cerny, Eunjee Song}}

\maketitle
\begin{abstract}

Cloud-native systems represent a significant leap in constructing scalable, large systems, employing microservice architecture as a key element in developing distributed systems through self-contained components. However, the decentralized nature of these systems, characterized by separate source codes and deployments, introduces challenges in assessing system qualities. Microservice-based systems, with their inherent complexity and the need for coordinated changes across multiple microservices, lack established best practices and guidelines, leading to difficulties in constructing and comprehending the holistic system view. This gap can result in performance degradation and increased maintenance costs, potentially requiring system refactoring. The main goal of this project is to offer maintainability assurance for microservice practitioners. It introduces an automated assessment framework tailored to microservice architecture, enhancing practitioners’ understanding and analytical capabilities of the multiple system perspectives. The framework addresses various granularity levels, from artifacts to constructing holistic views of static and dynamic system characteristics. It integrates diverse perspectives, encompassing human-centric elements like architectural visualization and automated evaluations, including coupling detection, testing coverage measurement, and semantic clone identification. Validation studies involving practitioners demonstrate the framework’s effectiveness in addressing diverse quality and maintainability issues, revealing insights not apparent when analyzing individual microservices in isolation.

\end{abstract}



\section{Introduction}

Cloud-native design emerged as the mainstream direction for decentralized systems seeking scalability and high performance under heavy workloads. 
The adoption of microservice architecture for such direction is crucial for fully harnessing the advantages of cloud computing~\cite{Cerny:2018:CUM:3183628.3183631}. Microservices, with their self-contained nature, enhance management and deployment efficiency, making them highly favored in enterprise software systems. The microservice architecture, once considered cutting-edge, has now become the standard, with industry giants like Amazon and Netflix leading the way. These paradigms have profoundly transformed application development, emphasizing scalability and efficiency.


\subsubsection*{Problem Statement} The problem statement revolves around system maintainability. In particular, it considers challenges stemming from comprehension of the overall system, attributed to poor design decisions and heightened system complexity. Software systems undergo transformations throughout their lifecycle, whether to incorporate new features, adapt to different environments, address bugs, and more. However, these changes often pose significant challenges. As highlighted by Bass et al.~\cite{bass2003software}, the software development community is grappling with the realization that about 80\% of the total cost of a typical software system is incurred after its initial deployment to modify and maintain the system. Consequently, a significant portion of the systems that developers engage with is in this post-deployment phase. Many software practitioners work within the constraints of the existing architecture and codebase, with limited opportunities for new development.


Moreover, managing a complex architecture, characterized by its dynamic, distributed, and expansive nature, poses significant challenges. Independent decision-making by teams on parts of the system, coupled with the evolving nature of the system, can lead to ripple effects where changes in one part impact another, complicating management processes. Additionally, maintaining up-to-date documentation for numerous interconnected services, each with multiple dependencies, presents substantial challenges in management and maintenance. These efforts may result in downtime and necessitate a comprehensive overhaul of the entire system.





\subsubsection*{Research Objective} The objective is to introduce and implement an automated assessment framework for microservice systems. This objective is fueled by building comprehensive foundations especially with the absence of comprehensive guidelines governing microservices practices. This framework aims to enhance analytical abilities and reasoning in architectural design, identify early indicators of system design degradation, and offer testability measurements for the system. 


\subsubsection*{Paper Organization} The rest of this paper is organized as follows: Section II covers the background and objectives with research questions (RQs). Section III details the methodology and prior results. Section IV discusses the development of the maintainability framework. Section V outlines the research contributions and conclusion.






\section{Background and Objective} 






Microservices, foundational to cloud-native applications, are designed as small, interdependent services that decompose functionalities into modular, autonomous units, thereby enhancing agility and resource utilization~\cite{abdelfattah2023roadmap,aws_cloud_native}. These microservices interact with each other through network protocols and message passing mechanisms to fulfill various tasks. This architecture aims to improve system modularity, scalability, and maintainability.


Moreover, maintainability, as a crucial quality attribute, reflects the effectiveness and efficiency with which a product or system can be modified by its intended maintainers~\cite{bass2003software}. In line with the ISO 25010~\cite{ISO25010} standards outlining software quality requirements, maintainability encompasses fundamental quality dimensions intrinsic to both the system and its architecture, including modularity, reusability, analyzability, modifiability, and testability.


The main objective is to design and implement an automated assessment framework for the maintainability of microservice systems. To achieve this, several RQs must be explored, each contributing to the framework and supporting system maintainability.

\begin{center}
	\begin{rqbox}
		\textbf{RQ$_1$} \emph{What dependencies exist within microservices-based systems?}
    \end{rqbox}
\end{center}
\vspace{-0.5em}
In microservices-based systems, dependencies become distributed, heterogeneous, less visible, and harder to track. The challenge is compounded by the absence of comprehensive guidelines and catalogs for identifying and defining these dependencies. Such guidelines are crucial for enhancing system maintainability. \textit{This RQ aims} to develop a dependency taxonomy to offer a comprehensive understanding of these dependencies, their impact, and their categorization.

\begin{center}
	\begin{rqbox}
		\textbf{RQ$_2$} \emph{How to construct system-centric views of microservice architecture?}
    \end{rqbox}
\end{center}
\vspace{-0.5em}
The decentralized nature of microservices and their dependencies presents a challenge wherein alterations in one part of the architecture can yield widespread effects, both explicitly and implicitly. Teams frequently enact these changes without possessing a comprehensive view of the entire system. Moreover, the isolated nature of teams, each operating within separate sandboxes, fosters independent decision-making, potentially influencing the overall architecture. \textit{This RQ aims} to develop holistic and centric views of the system architecture that encompass various perspectives of dependencies.

\begin{center}
	\begin{rqbox}
		\textbf{RQ$_3$} \emph{How to Implement automatic testability and quality assessment methodologies on the system?}
    \end{rqbox}
\end{center}
\vspace{-0.5em}
A critical challenge facing microservice systems is the absence of tailored quality assessment techniques. Existing methodologies often fail to address the unique characteristics and complexities inherent in microservice architectures \cite{ghani2019microservice,jiang2022efficient}. This deficiency hampers the ability of development teams to effectively evaluate the maintainability of their microservice-based applications, thus hindering their overall quality assurance efforts. \textit{This RQ aims} to employ holistic views to formulate and automate assessment metrics and techniques.  It seeks to adopt a holistic approach that considers various facets of the system architecture and its operation to create robust and efficient evaluation methodologies.

\begin{center}
	\begin{rqbox}
		\textbf{RQ$_4$} \emph{How can a tailored visualization methodology be designed to effectively communicate the constructed views to practitioners?}
\end{rqbox}
\end{center}
\vspace{-0.5em}
The extensive information encompassed within microservice architecture and its holistic views presents a challenge in effectively conveying this wealth of data to practitioners. Furthermore, the absence of visualization tools specifically tailored for microservices views raises questions about the adequacy of traditional visualization techniques in capturing microservice perspectives. \textit{This RQ aims} to design and validate a visualization technique specifically tailored for depicting representations of microservices. It aims to develop a visualization approach that effectively captures the intricacies and dependencies inherent in microservice architectures, providing practitioners with intuitive and insightful visualizations to aid in their understanding and reasoning processes.

\section{Methodology and Prior Results}

Different research questions necessitate distinct methodologies to address them. Employing diverse methodologies enhances the breadth of outcomes and validates their robustness.

Regarding \textbf{RQ$_1$}, it embarks on the ambitious endeavor of conducting a comprehensive study to grasp the foundational issues stemming from the core concept of dependencies. This study is centered on the creation of a varied dependency taxonomy, which amalgamates insights from existing literature, tackles challenges to comprehend existing dependencies, and draws upon various expertise to elucidate multifaceted dependency dimensions. It distinguishes between root causes and symptomatic occurrences, thus reducing ambiguity during identification. \textit{To tackle this RQ}, two methodologies are utilized: Systematic Literature Review (SLR)\cite{kitchenham2013systematic} and Open and Axial coding\cite{khandkar2009open,kendall1999axial}. The SLR collects literature evidence to understand current microservice dependencies. Then, Open and Axial coding validate and categorize these dependencies with expert collaboration. These methodologies are evaluated for their systematic approach and efforts to reduce bias by involving external practitioners.


Prior results pertaining to this RQ include a discussion paper on the correlation between dependencies and system maintainability, as referenced in~\cite{closer24-dependency}. Furthermore, the dependency taxonomy (under review) reveals both implicit and explicit types of dependencies. Explicit dependencies occur when one microservice directly calls another via a REST interface~\cite{masse2011rest}. Similarly, data dependencies arise when multiple data entities are shared between microservices within a bounded context. Implicit dependencies arise through semantic clones, where multiple microservices share similar business logic.

For \textbf{RQ$_2$}, the aim is to leverage the dependency taxonomy and constructed guidelines to build system-centric views of microservice-based systems. This involves creating multidimensional perspectives (viewpoints) based on various dependency artifacts, which form the cornerstone of the research endeavor, utilizing the established guidelines to formulate subsequent research questions and provide practitioners with a comprehensive system view. \textit{To address this RQ}, the Software Architecture Reconstruction (SAR)~\cite{guo1999software} methodology is employed to automatically generate these comprehensive views. Additionally, a review methodology is utilized to develop a roadmap for generating different representations from the system artifacts. A review has been conducted to introduce this roadmap~\cite{abdelfattah2023roadmap} for constructing the necessary viewpoints. Both static and dynamic analyses are employed to capture the multiple dimensions of the system. Prototypes and case studies are then developed to assess the effectiveness of the method and the completeness of the generated viewpoints.

\begin{figure}[h!]
\vspace{-1em}
        \centering
        \includegraphics[width=\linewidth]{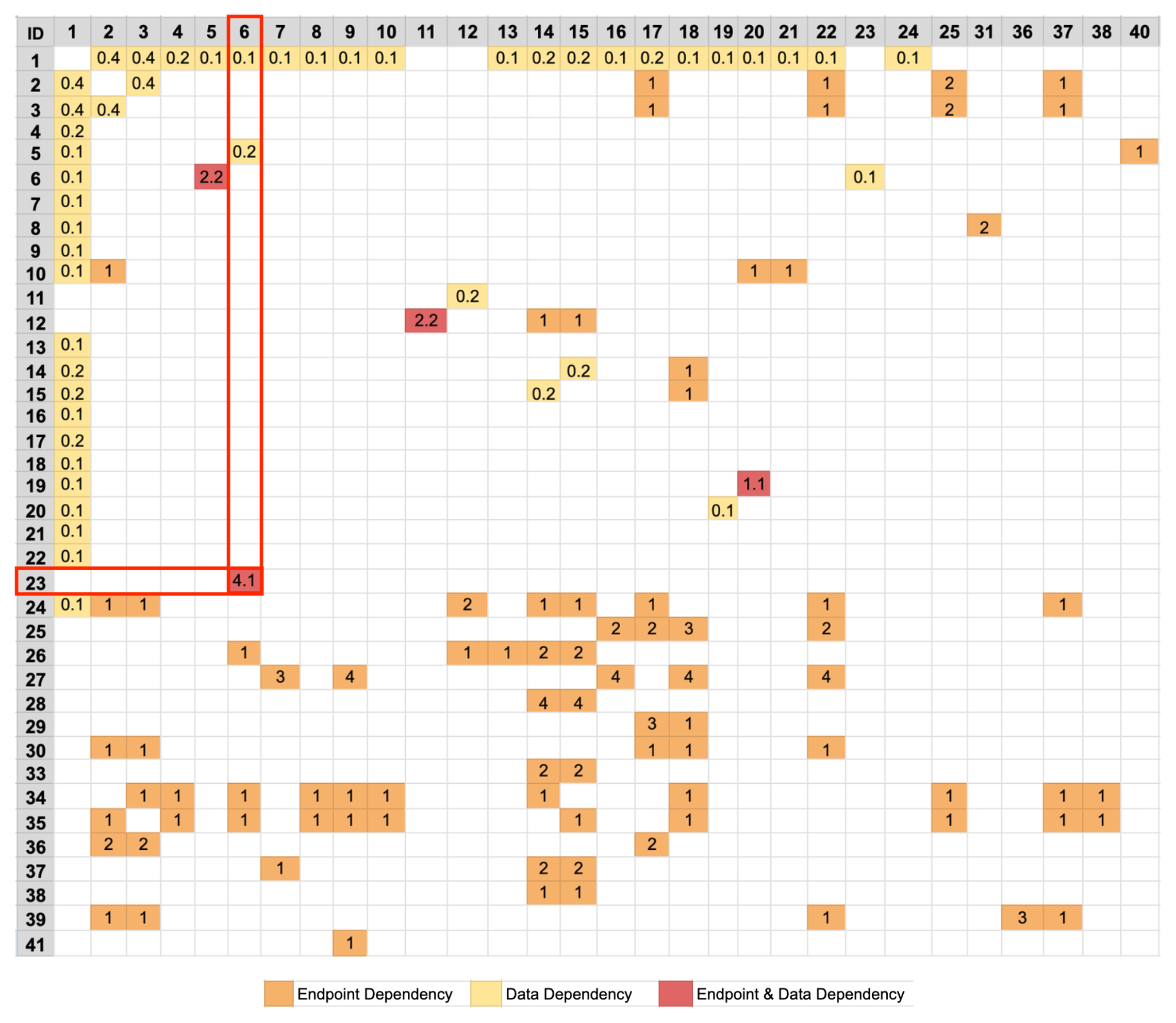}
        \vspace{-1em}
       \caption{Service Dependency Matrix (SDM). This matrix is sourced from Abdelfattah et al. (2023)~\cite{abdelfattah2023microservice}.}
      \vspace{-0.7em}
       \label{fig:sdm}
    \end{figure}

Prior results have been derived from endpoint dependencies and data dependencies extracted using static analysis techniques. These results are presented in dependency matrix representations as outlined in~\cite{abdelfattah2023microservice}. These matrices offer a central viewpoint illustrating the behavior of dependencies across dimensions, such as the Service Dependency Matrix (SDM) depicted in~\fig{sdm} (generated from the TrainTicket benchmark~\cite{trainticketgit}). For example, considering the cell at position (row: 23, column: 6) in the SDM, with a value of 4.1, it signifies that the \texttt{ts-admin-user-service} microservice (ID 23) has made four calls to the \texttt{ts-user-service} microservice (ID 6), with one shared entity (\texttt{UserDto}) between them. These dependencies underscore their interconnected nature and the significance of their mutual interaction.

For \textbf{RQ$_3$}, two main tasks are addressed: firstly, creating a comprehensive catalog of microservice bad practices to establish benchmarks for assessing architecture maintainability; secondly, leveraging the developed holistic views to devise and automate assessment metrics and techniques.

Transferring design patterns across various software architectures is challenging, as not all patterns fit universally, leading to anti-patterns. To gauge maintainability, current systems are assessed for anti-patterns, often using tools like SonarQube~\cite{SonarQube}. However, SonarQube evaluates individual applications and cannot identify microservice smells or detect anti-patterns across codebases. This highlights the need to explore and identify anti-patterns specific to microservices for effective system maintenance. \textit{To achieve this}, this research utilizes the Triatry systematic literature review~\cite{kitchenham2010systematic} in conjunction with the Open and Axial coding methodologies to construct a cohesive and exhaustive catalog focused on microservice ecosystem anti-patterns. Additionally, it leverages the developed viewpoints and integrates dynamic data to assess the system by analyzing runtime behavior and computing assessment metrics such as endpoint test coverage metrics.

Prior results introduced a comprehensive anti-pattern catalog in~\cite{cerny2023catalog}, comprising 5 main categories and 58 individual anti-patterns. This catalog engages experts from the field to ensure consistency and merge similar anti-patterns published under different names or with slightly varied definitions. It lays the groundwork for developing assessment techniques to identify and address bad practices within microservice systems. Regarding assessment metrics, this research has delivered End-to-End (E2E) and API test coverage metric calculations, as detailed in~\cite{electronics13101913,abdelfattah2023end}. It presents three metrics for endpoint test coverage calculation in microservice systems. Moreover, it encompasses additional metrics aimed at gauging the impact of evolutionary changes across various components of the system.

For \textbf{RQ$_4$}, the focus lies on determining the suitable design approach to effectively convey the complex information extracted from the microservice ecosystem. The challenge lies in visualizing the holistic view in a more engaging and interactive manner, rather than relying solely on static dependency matrices. This prompts the question of whether conventional methodologies suffice for representing this information or if a tailored visualization methodology should be introduced. \textit{To tackle this RQ}, we initiated by conducting a thorough literature review to assess the state of the art in visualization within this domain. Subsequently, we devised and developed a visualization methodology, followed by a series of controlled experiments involving practitioners to evaluate the understandability of microservice dependency behavior through both conventional and tailored visualization approaches.

Prior results performed multiple reviews addressing visualization in the literature~\cite{cerny2022microservice,burgess2022visualizing}, which highlighted a shortage in microservice visualization methods, necessitating a tailored design to encompass its unique perspectives. We then presented the design outline of the visualization methodology experiment in~\cite{abdelfattah2022microservices}, which involved implementing a tool named Microvision utilizing Augmented Reality (AR) medium to address the challenge of limited rendering space in traditional visualization, as detailed in~\cite{cerny2022microvision}. For assessment purposes, we conducted controlled experiments comparing Microvision with conventional 2D-graph-based visualization tools, with the protocol and results analysis published in~\cite{abdelfattah2023comparing}. The 2D tool uses rectangular boxes and arrows to present microservice dependency graphs, similar to commercial and open-source tools. In contrast, the Microvision tool renders a 3D model with cubes and line connectors, displaying call information via popups. Both visualizations convey the same information but offer distinct display and interaction features for different needs. This experiment involved systems of two sizes (small and large) and participants with varying expertise (novice and expert). The findings showed that AR effectively aids in understanding microservice architecture, enabling novice practitioners to grasp system dependencies like experienced users. While 2D tools clearly visualized dependencies in small systems, Microvision outperformed in visualizing large systems with scalability in mind.



\begin{figure}[h!]
\vspace{-1em}
        \centering
        \includegraphics[width=0.95\linewidth]{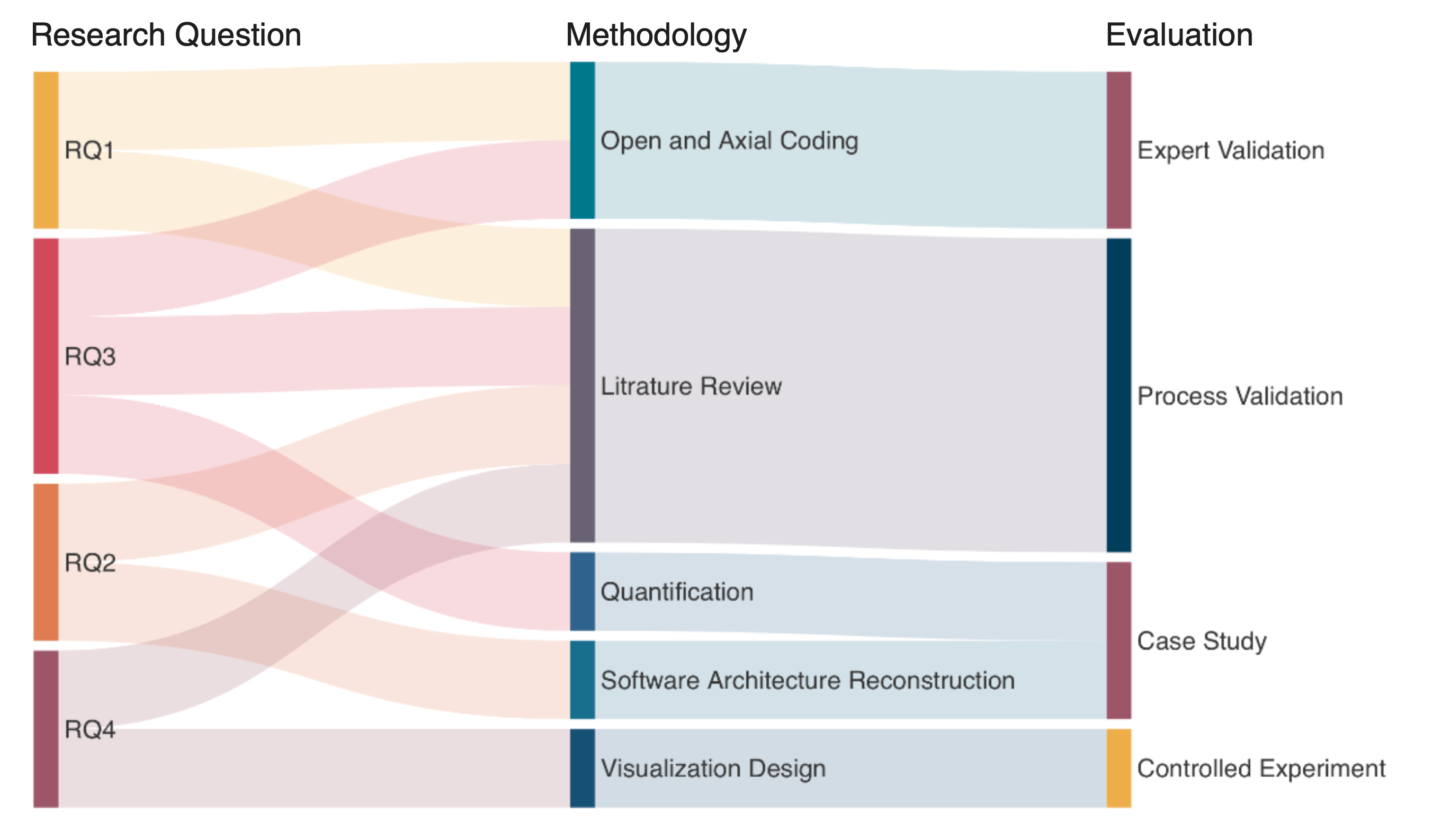}
        \vspace{-0.7em}
       \caption{Methodologies aligned with research questions.}
      \vspace{-0.5em}
       \label{fig:research-methodology}
    \end{figure}

As depicted in~\fig{research-methodology}, each research question employs a distinct literature review methodology tailored to its specific tasks. RQ$_1$ and RQ$_3$ utilize the Open and Axial coding methodology to refine the data extracted from the literature review. Regarding evaluation methodologies, this study assesses each method based on its inherent nature. The literature review undergoes validation through adherence to process guidelines, while the Open and Axial coding process is validated through collaboration with experts. Case studies are employed to validate the Software SAR process and quantification metrics. Additionally, controlled experiments are utilized to evaluate the effectiveness of the visualization design methodology.

\section{Framework Construction}


These RQs' perspectives form the basis for developing an assessment framework to allow analyzers to operate based on extracted components and constructed holistic views.

\begin{figure}[h]
    \vspace{-1.3em}
    \centering
    \includegraphics[width=0.9\linewidth]{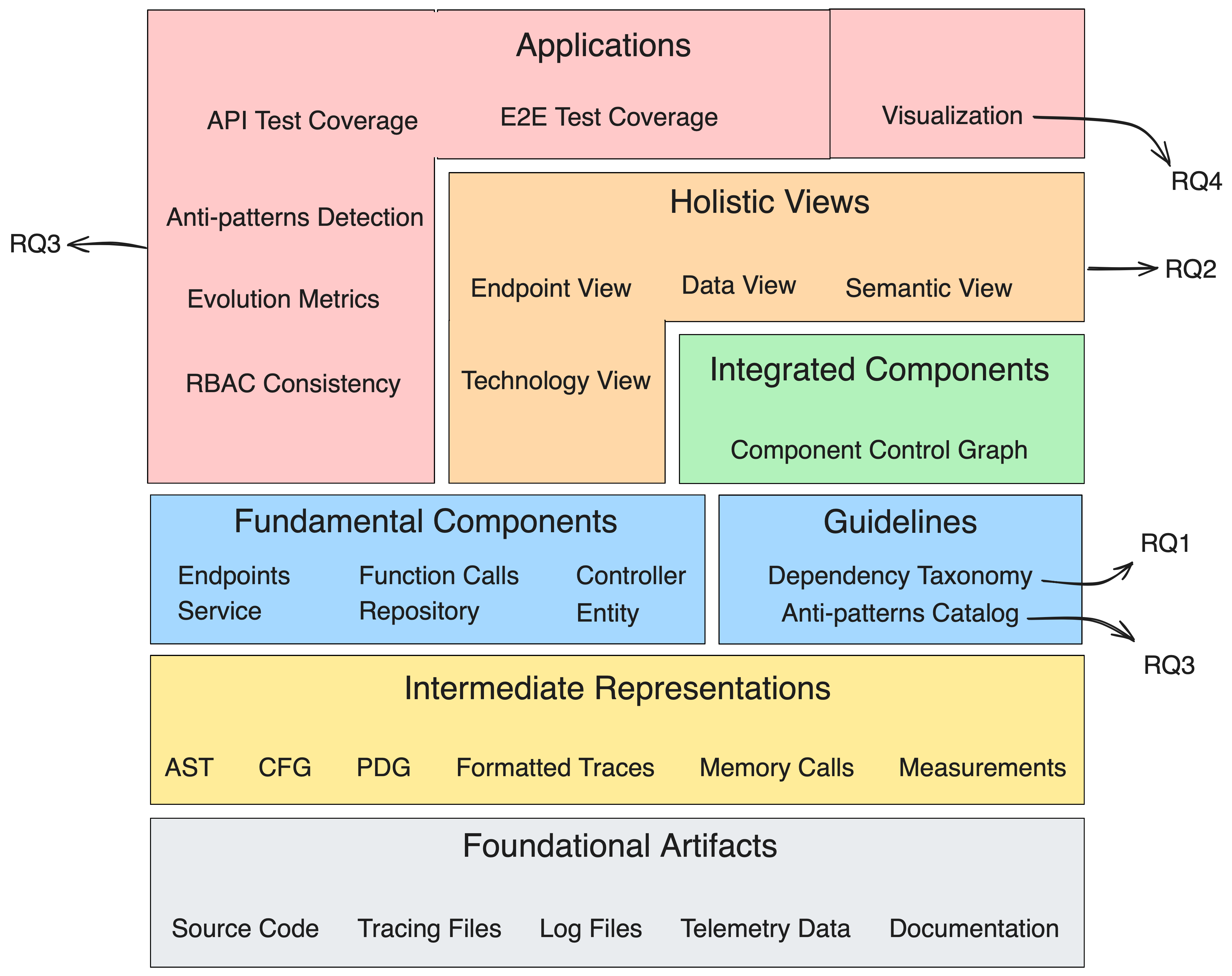}
    \caption{The proposed Microservice Assessment Framework}
    \label{fig:framework}
    \vspace{-1.5em}
    \end{figure}
    
This framework, as depicted in~\fig{framework} is centered around the construction of the system holistic view and derive automated application to provide assessment for the system quality perspectives. This framework considers various system granularities and perspectives, combining human-centric aspects like architectural visualization with automated evaluation facets. This framework adheres to the principles outlined in layers-based guidelines~\cite{bass2003software}, which promote the use of layers to offer guidance and flexibility in identifying architectural aspects and determining the views to be generated. This six-layered framework originates from the microservice system artifacts and culminates at the application layer, the epicenter of assessment objectives. The application layer (both quality assessment perspective (RQ$_3$) and visualization perspective (RQ$_4$)) harmonizes with one of three granularity component layers. At the fundamental components layer, the system components are extracted from the constructed intermediate representation, an outcome of the intricate static and dynamic analysis extraction processes. Above it lies the integrated components~\cite{abdelfattah2023detecting} layer, a flow of interconnected components within individual microservices. Contrarily, the holistic views (RQ$_2$) layer orchestrates the interconnected components and flows across multiple microservices, constructing a comprehensive view of the entire system. Moreover, the guidelines (dependency taxonomy (RQ$_1$) and anti-pattern catalog (RQ$_3$)) serve as valuable resources to augment the understanding and application of microservices practices. This concerted effort to compile such essential references directly fuels the fundamental component of this framework.


The introduced framework provides flexibility to the application layer, allowing diverse applications to assess different system perspectives. The outcomes demonstrate that this framework effectively addresses issues through a holistic viewpoint, uncovering insights that might be missed when analyzing individual microservices in isolation.

\section{Contribution and Conclusion}




This work introduces a comprehensive framework for microservices systems, enhancing system maintainability and analytical capabilities. The framework offers techniques for effective system modifications, ensuring testability, and improving modularity and reusability to maintain manageability.

In addition to practical applications, this work provides valuable resources to the community. These include multiple publications detailing the framework's construction, comprehensive catalogs for anti-patterns and dependency taxonomy, datasets with dependency measurements, anti-pattern definitions, testing suites, and a visualization tool for holistic service views, integrated as a plugin with development tools.

The framework paves the way for future research, such as exploring new holistic perspectives from technological and operational standpoints, broader applications like interactive documentation, and incorporating evolutionary aspects through co-change analysis.

\bibliographystyle{IEEEtran}
\bibliography{references}

\end{document}